\documentclass[aps,preprint,prb,amsmath,showpacs]{revtex4}
\usepackage{graphicx}
\usepackage{bm}

\begin{document}
  
\title{Elastic Properties of Nanowires}

\author{Alexandre F. da Fonseca}
\email[Eletronic mail: ]{afonseca@if.usp.br}
\author{C. P. Malta}
\email[Eletronic mail: ]{coraci@if.usp.br}

\affiliation{Instituto de F\'{\i}sica, Universidade de S\~ao
Paulo, USP, Rua do Mat\~ao Travessa R 187, CEP 05508-900, S\~ao Paulo,
SP, Brazil}

\author{Douglas S. Galv\~ao}
\email[Eletronic mail: ]{galvao@ifi.unicamp.br}

\affiliation{Instituto de F\'{\i}sica `Gleb Wataghin',
Universidade Estadual de Campinas, UNICAMP CEP 13083-970, Campinas,
SP, Brazil}

\begin{abstract} 

We present a model to study Young's modulus and Poisson's ratio of the
composite material of amorphous nanowires. It is an extension of the
model derived by two of us $[$Da Fonseca and Galv\~{a}o,
Phys. Rev. Lett. {\bf 92}, 175502 (2004)$]$ to study the elastic
properties of amorphous nanosprings. The model is based on twisting
and tensioning a straight nanowire and we propose an experimental
setup to obtain the elastic parameters of the nanowire. We used the
Kirchhoff rod model to obtain the expressions for the elastic
constants of the nanowire.

\end{abstract} 
 
  

\maketitle

\section{Introduction}  

The growing interest in the field of nanoscience comes, in part, from
the discovery of one-dimensional carbon nanotubes~\cite{iijima} and
its special physical properties as, for example, the exceptionally
high Young's modulus ($\sim1$ TPa). It has attracted the attention of
the scientific community to the wide potential of the nanostructures
to diverse scientific and technological
applications.~\cite{baugh,ede,wo1,gu1,liu,sam}

Among the different types of nanostructures, carbon nanotubes,
nanowires and nanosprings present special mechanical properties. In
order to understand their macroscopic mechanical behavior it is
necessary to study their elastic properties at nanoscale. Scanning
transmission microscopy (STM), atomic force microscopy (AFM) and
transmission electron microscopy (TEM) have been used to manipulate
and characterize the properties of individual
nanostructures.~\cite{wang1,ugarte,li,wang2}

In this paper, we present a model for the study of the elastic
properties of amorphous straight nanowires. We propose an experimental
setup that can be used to measure the Young's modulus and the
Poisson's ratio of the composite material of an amorphous nanowire in
two situations. 

Our proposal is based on two situations involving a twisted nanowire.
The first situation is a static configuration in which the nanowire is
maintained twisted by a given torque. From the equations of
equilibrium for a filament in this situation, we obtain an expression
for the elastic parameters of the nanowire. The second situation
involves the twisting and tensioning of a nanowire. In order to obtain
a second expression for the elastic parameters of the nanowire, we use
the well known fact that a twisted rod becomes unstable if its twist
density reaches a given critical value.

The {\it Kirchhoff rod model}~\cite{kirchhoff} is used to derive
simple equations for the elastic constants of the composite material
of an amorphous nanowire in the two above mentioned configurations. These
equations involve the geometry of the twisted nanowire and the forces
and torques that hold the nanowire deformed. Since we are considering
amorphous nanowires, a continuous mechanical model is appropriate to
study the elastic properties of these systems. The present work is a direct
extension of a previous work where the Kirchhoff rod model was used to
derive two expressions for measuring the Young's modulus and Poisson's
ratio of amorphous nanosprings.~\cite{fonseca5}

\section{The mechanical rod model}

The Kirchhoff model is a theory for small deformations of inextensible
thin elastic rods. It has been extensively used to study the statics
and the dynamics of different kinds of continuous filaments with
applications ranging from
Biology~\cite{olson,coleman,fonseca1,fonseca2,fonseca3,alain1,alain2}
to Engineering.~\cite{coyne,sun}

The Kirchhoff equations are derived from the application of Newton's
laws of mechanics to a thin elastic rod in the approximation of small
curvature.~\cite{dill} An additional condition is needed to solve the
set of differential equations and it comes from the constitutive
relation of linear elasticity theory that relates the moments to
strains.~\cite{dill} These equations contain the forces and torques,
plus a triad of vectors describing the deformations of the rod. For a
circular homogeneous rod, they are given by
\begin{subequations}
\label{kir}
\begin{eqnarray}
&&{\bf F}''  =  \rho A \ddot{{\bf d}}_3  \; , \label{a}  \\
&&{\bf M}' + {\bf d}_{3}\times {\bf F}  =  
\rho I \left({\bf d}_1\times\ddot{{\bf d}}_1+
{\bf d}_2\times\ddot{{\bf d}}_2 \right)  \; , \label{b}  \\
&&{\bf M}  =  E\,I\,k_1{\bf{d}}_1 + E\,I\,k_2{\bf{d}}_2 + 
\frac{E\,I\,k_3}{1+\sigma}{\bf{d}}_3 \; , \label{c}  
\end{eqnarray}
\end{subequations}
where ${\bf F}$ and ${\bf M}$ are the total force and torque across
the cross-sections of the rod, respectively. ${\bf{d}}_i$ with
$i=1,2,3$ form a triad of unitary vectors defined in each
cross-section where ${\bf{d}}_3$ is the vector tangent to the axis of
the rod and ${\bf{d}}_1$ and ${\bf{d}}_2$ lie on the plane of the
cross-section along, for example, its principal axes. $E$ and $\sigma$
are the Young's modulus and the Poisson's ratio of the composite
material of the rod. $\rho$ is the mass density of the rod, and $I$ is
the moment of inertia of the cross section, here considered as being
circular of area $A$. If $r$ is the radius of the cross section,
\begin{equation}
\label{I}
I=\frac{\pi r^4}{4} \; .
\end{equation}
In Eqs. (\ref{kir}), the prime and the dots denote the derivative with
respect to the arc-length $s$ of the rod and to the time $t$,
respectively. $k_i$, $i=1,2,3$, are the components of the so-called
{\it twist vector}, ${\mathbf{k}}$, which defines the variation of the
director basis $\{{\bf d}_1,{\bf d}_2,{\bf d}_3\}$ with the arc-length
$s$ through the expression: ${\bf d}'_i={\bf k}\times{\bf d}_i$,
$i=1,2,3$. The Eq. (\ref{c}) is valid for an intrinsically straight
rod, i. e., a rod that is straight when free from the action of
stresses. That is the case of an amorphous straight nanowire.

\section{The experimental model} 

One of the advantages of our proposal is the fact that with an unique
experimental setup we can perform two measurements that, combined, can
give the elastic parameters of the composite material of the nanowire.

\begin{figure}[ht] 
  \begin{center}
  \includegraphics[height=60mm,width=95mm,clip]{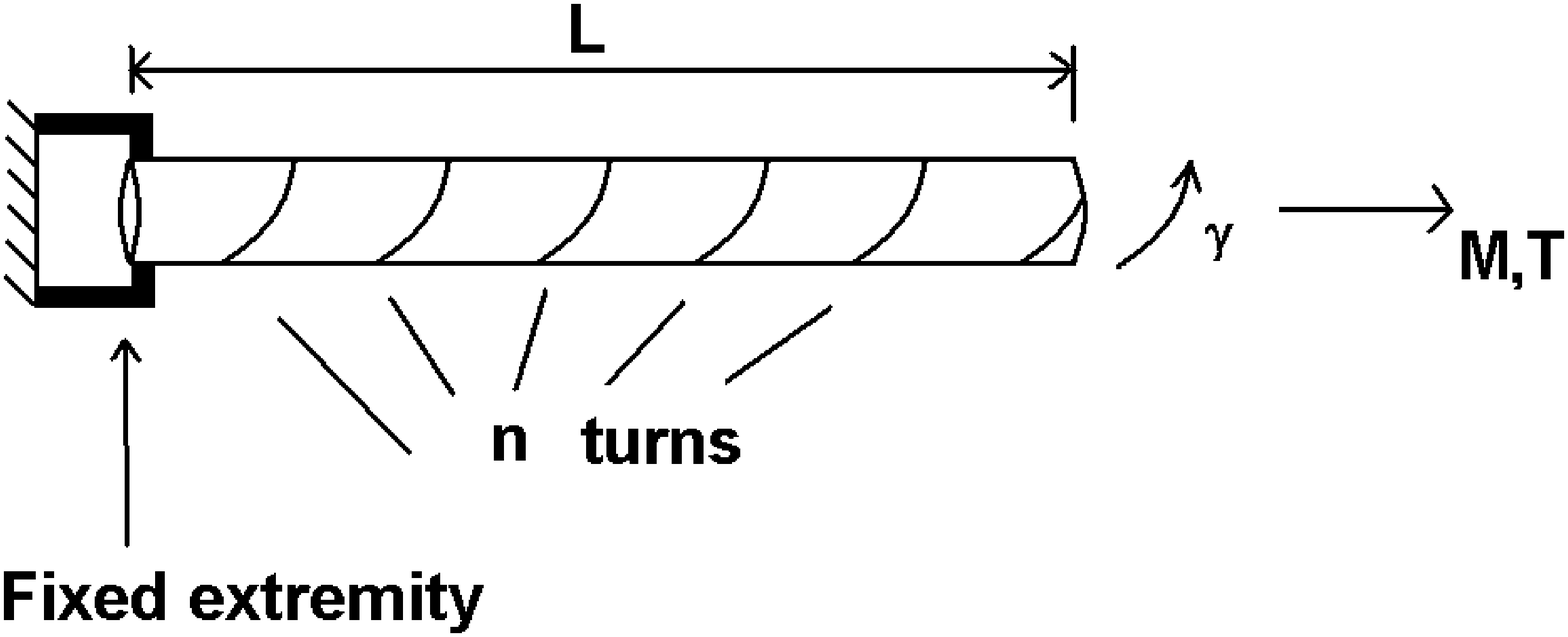} 
  \end{center}
  \caption{Outline of the experimental setup to measure the elastic 
  constants $E$ and $\sigma$. A torque ${\bf M}$ is applied along the 
  axis of the rod, creating a twist density $\gamma$. ${\bf T}$ is the 
  force of tension.}
\label{fig1}
\end{figure}

Fig. \ref{fig1} displays the proposed scheme to measure the
Young's modulus and the Poisson's ratio of the nanowire. Two
experiments can be performed with the same experimental setup. 

The first experiment is based on a static situation in which the
nanowire is twisted a given number $n$ of turns through the
application of a torque ${\bf M}$ along the direction of the axis of
the nanowire. The second experiment is based on the application and
control of a force of tension ${\bf T}$ on a twisted nanowire.

In order to derive the expression to be used in the first experiment,
we have to obtain an equilibrium solution of the Kirchhoff equations,
{\it i. e.}, a solution for the case where the time derivatives are
eliminated from the Eqs. (\ref{kir}). The equilibrium solution
corresponding to a twisted straight rod is given by
\begin{subequations}
\label{ttwist}
\begin{eqnarray}
&&{\bf k}=\gamma\,{\bf d}_3 \; , \label{t1} \\
&&{\bf F}=T\,{\bf d}_3 \; , \label{t2} \\
&&{\bf M}=(1+\sigma)^{-1}EI\gamma\,{\bf d}_3 \; , \label{t3}
\end{eqnarray}
\end{subequations}
where $\gamma$ is the twist density of the rod that is defined by:
\begin{equation}
\label{gama}
\gamma=\frac{n}{L} \; ,
\end{equation}
where $n$ is the number of turns produced by the application of the
torque ${\bf M}$, and $L$ is the total length of the nanowire. $T$ is
a force of tension or compression applied along the axis of the rod
and it is a free parameter in the static situation. The value of $T$
will be important in the second experiment which involves the
condition for the stability of the equilibrium solution given by
Eqs. (\ref{ttwist}). 

We rewrite the Eq. (\ref{t3}) in order to obtain an expression for the
elastic constants in terms of the applied torque and the twist
density:
\begin{equation}
\label{sigE}
\frac{E}{\sigma+1}=\frac{4\|{\mathbf{M}}\|}{\pi r^4 \gamma} \; ,
\end{equation}
where we have used the Eq. (\ref{I}) to substitute the expression for
the moment of inertia, $I$, in terms of the cross section radius, $r$.

In the first experiment, we apply a given torque ${\bf M}$ to twist a
nanowire a given number $n$ of turns. By measuring the value of
$\|{\mathbf{M}}\|$ and calculating the resultant twist density
$\gamma$ in a nanowire of cross-section radius $r$, we obtain the
ratio $E/(\sigma+1)$ of the nanowire using Eq. (\ref{sigE}). The next
experiment will provide the other relation between $\sigma$ and $E$.

In Eq. (\ref{gama}), $n$ does not have to be an integer. Any fraction
(for instance, a torsion of about 30 degrees) of a turn can be
considered in the experiment because the twist density is a real
number. Our proposal is based only on elastic deformations of the
nanowire in the linear regime. Therefore, the plastic regime can be
avoided by making a small number of turns or a fraction of a turn on
the nanowire. It is important to measure the magnitude of the applied
torque that produces a given twist density. It is also important to
maintain one of the extremities held fixed in order to prevent the
relaxation of the nanowire.

In order to discuss the second experiment, let us briefly introduce the
dynamical method of stability analysis developed by Goriely and
Tabor,~\cite{alain3,alain4} to study the stability of equilibrium
solutions of the Kirchhoff equations for twisted rods. Their method is
based on an expansion of the director basis of the deformed
configuration around the director basis of the stationary solution,
and they use the Eqs. (\ref{kir}) to obtain different sets of equations for
each power of a given pertubative parameter $\epsilon$. The linear
analysis can be performed studying the equations at order
$O(\epsilon)$. Nonlinear analysis can be performed using the equations
for higher orders of $\epsilon$. From the linear and nonlinear
analysis of a twisted straight rod, Goriely and Tabor~\cite{alain4}
obtained the following expression for the twist density $\gamma_C$
above which a finite rod becomes unstable:
\begin{equation}
\label{critical}
\gamma_C=(1+\sigma)\sqrt{\frac{1}{L^2}+
\frac{4T}{EI}} \; .
\end{equation}

We can see now that the value of the tension $T$ is very important for
the stability of a twisted rod. High values of $T$ leads to high
values for $\gamma_C$, so that by increasing the tension on a twisted
rod, we can prevent it from buckling due to instability. In order to
define the second experiment, we rewrite the Eq. (\ref{critical}) to
obtain an expression for the critical value, $T_C$, of the force of
tension, for which a twisted rod, with fixed twist density $\gamma$,
becomes unstable:
\begin{equation}
\label{T_C}
T_C=\frac{EI}{4}\left(\frac{\gamma^2}{(1+\sigma)^2}-
\frac{1}{L^2}\right) \; .
\end{equation}

The second experiment is also based on twisting a nanowire but here we
apply and control the magnitude of the force of tension on it. The
scheme is shown in Fig. \ref{fig1}. In order to obtain the critical
value, $T_C$, of the force of tension, we must perform the following
steps: 1) we apply a force of tension $T$ along the axis of the
nanowire maintaining it tensioned; 2) a torque is applied along the
axis of the nanowire so as to rotate the nanowire a number $n$ of
turns, thus producing a twist density $\gamma$, as in the previous
experiment. Keeping the torque constant fixes the value of $\gamma$,
and we do not need to measure the magnitude of the applied torque; 3)
the magnitude of $T$ is slowly decreased until the nanowire becomes
unstable and gets buckled. The value of $T$ such that the nanowire
becomes unstable is the critical value $T_C$ that can be used in the
eq. (\ref{T_C}) to obtain another relation between $E$ and $\sigma$.

Combining the results obtained using the Eqs. (\ref{sigE}) and
(\ref{T_C}), we obtain the values for the Young's modulus, $E$, and
the Poisson's ratio, $\sigma$, of the nanowire. 

\section{Discussion and conclusion}

The great potential for scientific and technological applications of
different types of nanostructures, has motivated a large amount of
work on the physical and chemical properties of
nanostructures. In future technological applications, any device
composed of one-dimensional nanostructures might take advantage of the
individual properties, specially those related to the elastic
responses to bending or twisting.

Manipulation of individual nanowires is a great challenge in
characterizing these systems. STM and AFM have been the dominant
approaches towards nanomanipulation.~\cite{wang1,li} Bending of
nanowires~\cite{wang1,wang2,dikin,chen1} and stretching of
nanocoils~\cite{chen2} have been reported in the literature but there
is no report about the twisting nanowire experiments. Nakatani and
Kitagawa~\cite{naka} recently presented an atomistic study of twisting
a nanowire using molecular dynamics. We hope that our proposal can
motivate experimentalists to develop methods for twisting individual
nanowires.

An important advantage of our proposed model is the fact that with an
unique experimental setup, we can make two experiments to obtain
directly the Young's modulus and the Poisson's ratio of the
nanowire. Since an amorphous nanowire is modeled as a continuous rod,
these two elastic constants completely characterize~\cite{landau} the
elastic behavior of the nanowire.  

If we wish to obtain the shear modulus $\mu$ of the composite material
of the nanowire, we can use the following relation:~\cite{landau}
\begin{equation}
\label{shear}
\mu=\frac{E}{2(\sigma+1)} \; .
\end{equation}
We can see that Eq. (\ref{sigE}) gives directly the expression for
$2\mu$, thus $\mu$ can be directly measured using the first scheme
proposed.

Our model can be used to test and confirm previous measurements of the
elastic parameters of an amorphous nanowire. For example, alternating
eletrical fields have been used to excite mechanical resonance in
SiO$_{2}$ nanowires~\cite{dikin} in order to measure the Young's
modulus of a nanowire having diameter $\sim$ 100nm and length over
10$\mu$m. They obtained a value close to 47 GPa for the Young's
modulus of these SiO$_{2}$ nanowires that is smaller than that for the
bulk amorphous SiO$_{2}$ ($\sim$ 72 GPa). The Young's modulus of
thinner SiO$_{2}$ (diameter $\sim$ 50nm) was measured using
TEM~\cite{wang1} and the obtained value (approximatelly 28 GPa) is
even smaller than that for the bulk case.

In other example, a reported increase of the Young's modulus of
different nanowires of diameters smaller than 70 nm has been explained
in terms of surface tension effects.~\cite{cue} Our proposed
experiments can be useful to test these effects since they are not
based on bending and, therefore, do not generate variations in the
surface area of the nanowire.

Conversely, our model can be used to estimate the torque required to
twisting a given nanowire if its Young's modulus and Poisson's ratio
are known. This can be very useful in technological applications where
individual nanowires can be subjected to forces and torques.

In summary, we have presented an experimental scheme to measure the
Young's modulus and the Poisson's ratio of the material that composes
an amorphous nanowire. Two schemes were proposed based on the twisting
of a nanowire. One scheme is based on a static configuration of a
twisted nanowire and the other is based on the conditions for a
twisted rod to become unstable. The Kirchhoff rod model was used to
obtain simple expressions relating the Young's modulus and the
Poisson's ratio of the material to the forces and torques that hold
the nanowire deformed. These two elastic constants completely
characterize the elastic properties of an individual amorphous
nanowire. It is, therefore, an important contribution for the
development of mechanical applications in nano-engineering.

\acknowledgments 
This work was partially supported by the Brazilian agencies FAPESP,
CNPq, IMMP/MCT, IN/MCT and FINEP.

\newpage


\noindent List of figure captions \\

FIGURE 1: Outline of the experimental setup to measure the
elastic constants $E$ and $\sigma$. A torque ${\bf M}$ is applied
along the axis of the rod, creating a twist density $\gamma$. ${\bf
T}$ is the force of tension. 

\newpage

\end{document}